# Nondetection sampling bias in marked presence-only data

Trevor J. Hefley[1], Andrew J. Tyre[2], David M. Baasch[3] & Erin E. Blankenship[4]

[1]Department of Statistics and School of Natural Resources, University of Nebraska–Lincoln, 234 Hardin Hall, 3310 Holdrege Street, Lincoln, Nebraska 68583
[2]School of Natural Resources, University of Nebraska–Lincoln, 416 Hardin Hall, 3310 Holdrege Street, Lincoln, Nebraska 68583
[3]Headwaters Corporation, 4111 4th Avenue, Suite 6, Kearney, Nebraska 68845
[4]Department of Statistics, University of Nebraska–Lincoln, 343B Hardin Hall North, 3310 Holdrege Street, Lincoln, Nebraska 68583

**Correspondence**
Trevor J. Hefley, Department of Statistics and School of Natural Resources, 234 Hardin Hall, 3310 Holdrege Street, University of Nebraska–Lincoln, Lincoln, NE 68583, USA.Tel: +402 472 4054; Fax: +402 472 2946;
E-mail: thefley@huskers.unl.edu

**Funding Information**
This research was supported by funding from the Platte River Recovery Implementation Program and the National Science Foundation Integrative Graduate Education and Research Traineeship (NSF-DGE-0903469).

Received: 20 August 2013; Revised: 21 October 2013; Accepted: 23 October 2013

doi: 10.1002/ece3.887

**Abstract**

1  Species distribution models (SDM) are tools used to determine environmental features that influence the geographic distribution of species' abundance and have been used to analyze presence-only records. Analysis of presence-only records may require correction for nondetection sampling bias to yield reliable conclusions. In addition, individuals of some species of animals may be highly aggregated and standard SDMs ignore environmental features that may influence aggregation behavior.
2  We contend that nondetection sampling bias can be treated as missing data. Statistical theory and corrective methods are well developed for missing data, but have been ignored in the literature on SDMs. We developed a marked inhomogeneous Poisson point process model that accounted for nondetection and aggregation behavior in animals and tested our methods on simulated data.
3  Correcting for nondetection sampling bias requires estimates of the probability of detection which must be obtained from auxiliary data, as presence-only data do not contain information about the detection mechanism. Weighted likelihood methods can be used to correct for nondetection if estimates of the probability of detection are available. We used an inhomogeneous Poisson point process model to model group abundance, a zero-truncated generalized linear model to model group size, and combined these two models to describe the distribution of abundance. Our methods performed well on simulated data when nondetection was accounted for and poorly when detection was ignored.
4  We recommend researchers consider the effects of nondetection sampling bias when modeling species distributions using presence-only data. If information about the detection process is available, we recommend researchers explore the effects of nondetection and, when warranted, correct the bias using our methods. We developed our methods to analyze opportunistic presence-only records of whooping cranes (*Grus americana*), but expect that our methods will be useful to ecologists analyzing opportunistic presence-only records of other species of animals.

## Introduction

A prerequisite to successful management and conservation of species is determining environmental and geographical features that influence the distribution of population abundance. Ecologists, statisticians, and computer scientists have developed and applied an impressive array of sampling methods and computational tools to estimate the distribution of abundance (Buckland and Elston 1993; Guisan and Zimmermann 2000; Guisan et al. 2002; Manly et al. 2002; Elith et al. 2006; Pearce and Boyce 2006; Phillips et al. 2006); however, rare or recently extinct species present a challenge because feasible sampling protocols produce few, if any, sightings of the species. An alternative approach involves documenting and analyzing opportunistic presence-only records. Opportunistic presence-only records often lack information on sampling effort and can consist of haphazard accounts of where a species occurred (e.g., museum records) or citizen reported sightings (Elith and Leathwick 2007; Van Strien et al. 2013). Opportunistic presence-only records are often analyzed using species distribution models (SDMs), but are not suitable to model the true distribution of population abundance if the detection and reporting of records are biased (Araújo and Guisan 2006; Pearce and Boyce 2006; Kéry 2011; Monk 2013; Yackulic et al. 2013). For example, a species may be detected and reported at a higher rate near roads or other areas that are easily accessible. Nondetection sampling bias that is affected by environmental and geographical features will bias estimates, predictions, and potentially conclusions derived from SDMs (Dorazio 2012; Monk 2013).

Recently, multiple authors have unified methods for analyzing presence-only data by showing that many previously developed methods (e.g., MAXENT, logistic regression) are approximating an inhomogeneous Poisson point process model (IPPM; Warton and Shepherd 2010; Aarts et al. 2012; Dorazio 2012; Fithian and Hastie 2013; Renner and Warton 2013; Warton and Aarts 2013; Hastie and Fithian 2013). Prior to our work, at least two limitations to using an IPPM to analyze presence-only data remained. First, nondetection sampling bias occurs when the probabilities of detection and reporting of the potential presence-only records are not constant across the landscape. Ignoring nondetection sampling bias can result in the estimation of an apparent species' distribution and interpreting IPPM parameters and predictions (e.g., heat maps) as if they represented the true species' distribution will result in potentially incorrect inferences (Kéry 2011; Dorazio 2012). Nondetection bias has received some attention recently (Rota et al. 2011; Dorazio 2012; Fithian and Hastie 2013; Kramer-Schadt et al. 2013; Monk 2013; Phillips et al. 2013; Yackulic et al. 2013), but methods to identify and potentially correct for the bias in SDMs, including the IPPM, were lacking. Here, we argue that nondetection sampling bias is equivalent to missing data for which a well-developed classification system exists to determine whether bias correction is required. Second, dependence between locations of individuals within a group results in correlation among points; one of the assumptions of the IPPM is that points are independent. Although there are many methods to model spatial dependencies of points, methods to model the extreme spatial dependence, for example, of a flock of birds, were lacking (Cressie 1993; Diggle 2003; Zipkin et al. 2012; Renner and Warton 2013). We demonstrate two extensions to the IPPM that (1) corrects for detection bias and (2) explicitly models group size. We tested our methods using simulated data sets that emulate data that an ecologist or statistician is likely to analyze. Our methods were explicitly developed to analyze opportunistic presence-only records of whooping cranes (Austin and Richert 2001); however, we envision that our methods will be useful to ecologists analyzing opportunistic presence-only records of other species of animals.

## Materials and Methods

### Species distribution model

The IPPM is appropriate to model the location of points that are independent after conditioning on the environmental and geographical covariates. If the locations of individuals are independent, then the IPPM is appropriate to model the distribution of individuals. Many species, however, occur in groups. If individuals are treated as unique points, at a minimum, the individuals (points) that belonged to a group are not independent. Methods to test for independence of groups (i.e., point interactions) are well developed, and many methods exist to explicitly model point interactions (e.g., area-interaction model; Cressie 1993; Diggle 2003; Renner and Warton 2013). We proceed assuming that individuals occur in independent groups and that group locations can be modeled with an IPPM; however, the analyst should verify this assumption (Diggle 2003; Renner and Warton 2013).

The IPPM is similar to a generalized linear model with a Poisson response distribution because environmental covariates influence the group intensity through the log link function. The linear predictor can be written as:

$$\log(\lambda_{gi}) = \alpha_0 + X_{gi}\alpha_{gi}, \qquad (1)$$

where the vector $\lambda_{gi}$ is the group intensities, $\alpha_0$ is the intercept, $X_{gi}$ is the design matrix of environmental covariates, and $\alpha_{gi}$ is the vector of environmental coefficients.

To estimate model parameters, the IPPM likelihood is required. The IPPM likelihood contains an integral that can be difficult or impossible to solve; therefore, numerical approximation is required. Many techniques have been developed to approximate the likelihood and obtain parameter estimates from the IPPM, and several of the methods are implemented in easily accessible software packages (Fithian and Hastie 2013).

Additional data associated with presence-only locations (e.g., group sizes) are known as marks (Cressie 1993; Diggle 2003). Marked IPPMs, for example, have been applied in forestry statistics to model the locations of trees and wood volumes (Stoyan and Penttinen 2000). We treat group sizes as marks and analyze the marks using a zero-truncated generalized linear model (GLM) assuming a truncated Poisson distribution. The zero-truncated GLM is similar to standard GLMs; however, the assumed response distribution is conditioned on the fact that only group sizes greater than zero can be reported for presence-only data (Zuur et al. 2009; Zipkin et al. 2012). Similar to the IPPM model, we model the expected group size using a linear predictor

$$\log(\lambda_{gs}) = \gamma_0 + X_{gs}\gamma_{gs}, \quad (2)$$

where the vector $\lambda_{gs}$ is the rate parameters of the zero-truncated Poisson distribution (i.e., unconditional expected group sizes), $\gamma_0$ is the intercept, $X_{gs}$ is the design matrix of environmental covariates and $\gamma_{gs}$ is the vector of environmental coefficients.

Modeling group sizes separately from group locations allows us to use different covariates in models of group intensities and group sizes. This flexibility is required to adequately model the distribution of abundance if environmental features influence group sizes. We note that the zero-truncated Poisson distribution may not be the best model of group sizes for all presence-only data; however, many zero-truncated distributions (e.g., zero-truncated negative binomial) exist. Models of sea duck group sizes from aerial surveys were explored by Zipkin et al. (2012), and their methods could also be applied to presence-only data.

To model intensities of abundance ($\lambda_{abundance}$), we multiplied the elements of group intensities by the unconditional expected group sizes:

$$\lambda_{abundnace} = \lambda_{gi} \times \lambda_{gs}. \quad (3)$$

Due to the exponential inverse link function, environmental coefficients that occurred in both the IPPM and zero-truncated GLM models can be summed to estimate the marginal effects of environmental covariates on intensity of abundance.

Although we have presented linear models for the IPPM and zero-truncated GLM, many less restrictive methods exist to estimate $\lambda_{gi}$ and $\lambda_{gs}$. For example, boosted regression trees or generalized additive models could also be used to estimate $\lambda_{gi}$ and $\lambda_{gs}$ (Guisan et al. 2002; Elith et al. 2008; Fithian and Hastie 2013).

## Correcting for nondetection

Sampling bias that results in nondetection of groups has the potential to bias parameter estimates and predictions from the IPPM, zero-truncated GLM or any SDMs that uses presence-only data (Dorazio 2012). The effect of nondetection (i.e., Bernoulli thinning of the point process) on parameter estimates and predictions from an IPPM depends on the covariates that affect the detection and intensity process (i.e., $\lambda_{gi}$). Although the effects of nondetection on the IPPM have been documented (Dorazio 2012), we chose to conceptualize the detection process as a missing data mechanism so we could provide a unified framework that applies to both group locations and group sizes (Little and Rubin 2002). Using the terminology of Rubin (1976), if detection and reporting of groups were perfect (i.e., $p_{det} = 1$; where $p_{det}$ is the vector of probabilities corresponding to each presence-only record), opportunistic records would consist of every possible location of the groups. With perfect detection, all parameters estimates from the IPPM would be asymptotically unbiased and identifiable. If detection is imperfect, but the covariates that influence the detection process are independent of the covariates that affect $\lambda_{gi}$, then the missing data are classified as missing completely at random (MCAR). In general, MCAR data are the best that can be obtained from any presence-only data collection process. If the nondetected presence-only data are MCAR, unbiased coefficients and relative intensities ($\lambda_{gi} = \lambda_{gi\,relative}e^{\alpha_o}$) are estimated with the IPPM assuming the model is correctly specified; however, an unbiased intercept parameter ($\alpha_0$) is unidentifiable (Dorazio 2012; Fithian and Hastie 2013). If the covariates that affect the detection process are correlated or share covariates with the covariates affecting $\lambda_{gi}$, the missing data mechanism results in nonignorable missing (NIM) data and the coefficients of the correlated or shared covariates estimated from the IPPM will be biased (Dorazio 2012). It should be emphasized that covariates affecting the probability of detection that are the same as or correlated with covariates affecting $\lambda_{gi}$, but are not included in the IPPM due to model misspecification (i.e., neglecting to include the covariate), will result in NIM data. In practice, it is difficult or impossible to know whether the model is correctly specified or whether the data are MCAR, therefore assuming that missing data mechanism results in NIM data is a conservative assumption. We present a decision tree to aid researchers in deciding when correcting for nondetection sampling bias is required for the IPPM model (Fig. 1).

The effect of nondetection on the analysis of group size marks is slightly different. Similar to the IPPM, if the covariates that affect detection are independent of the covariates that affect group size, then the missing data mechanism is MCAR, which is equivalent to a completely random sample of group sizes. If the detection process resulted in MCAR data for group size, all parameters ($\gamma_0$ and $\gamma_{gs}$) are identifiable and unbiased if detection is ignored. If, however, the covariates affecting detection are correlated with or the same as covariates affecting group size, the missing data are classified as missing at random (MAR). Under MAR, all parameters ($\gamma_0$ and $\gamma_{gs}$) are identifiable and unbiased if detection is ignored assuming the model of group size is specified correctly and contains the covariates that were correlated with or affected both nondetection and group size. Under the MAR mechanism, the detection process would result in less data from values of covariates that resulted in low detection, but unbiased parameters estimates (e.g., $\gamma_0$ and $\gamma_{gs}$) and predictions of $\lambda_{gs}$. For example, detection may be high close to developed areas, but large groups may tend to avoid these areas. In this case, more observations of large group sizes could be reported from areas that the larger groups tend to avoid, but analysis of the group size data does not result in biased estimates of the intercept ($\gamma_0$) or coefficients ($\gamma_{gs}$). Finally, if detection depends on group size after adjusting for the influence of covariates, the missing data mechanism is NIM, and parameters estimated would be biased. For example, if detection is greater for larger groups, then the parameters estimates from the zero-truncated GLM are biased and a correction for nondetection may be warranted. We present a decision tree to aid researchers in deciding when correcting for nondetection sampling bias is required for marks associated with presence-only locations (Fig. 2). Again, in practice, it is difficult or impossible to know whether the model is correctly specified or whether the missing data are MAR or MCAR, therefore assuming that missing data mechanism for the marks results in NIM data is likely a conservative assumption.

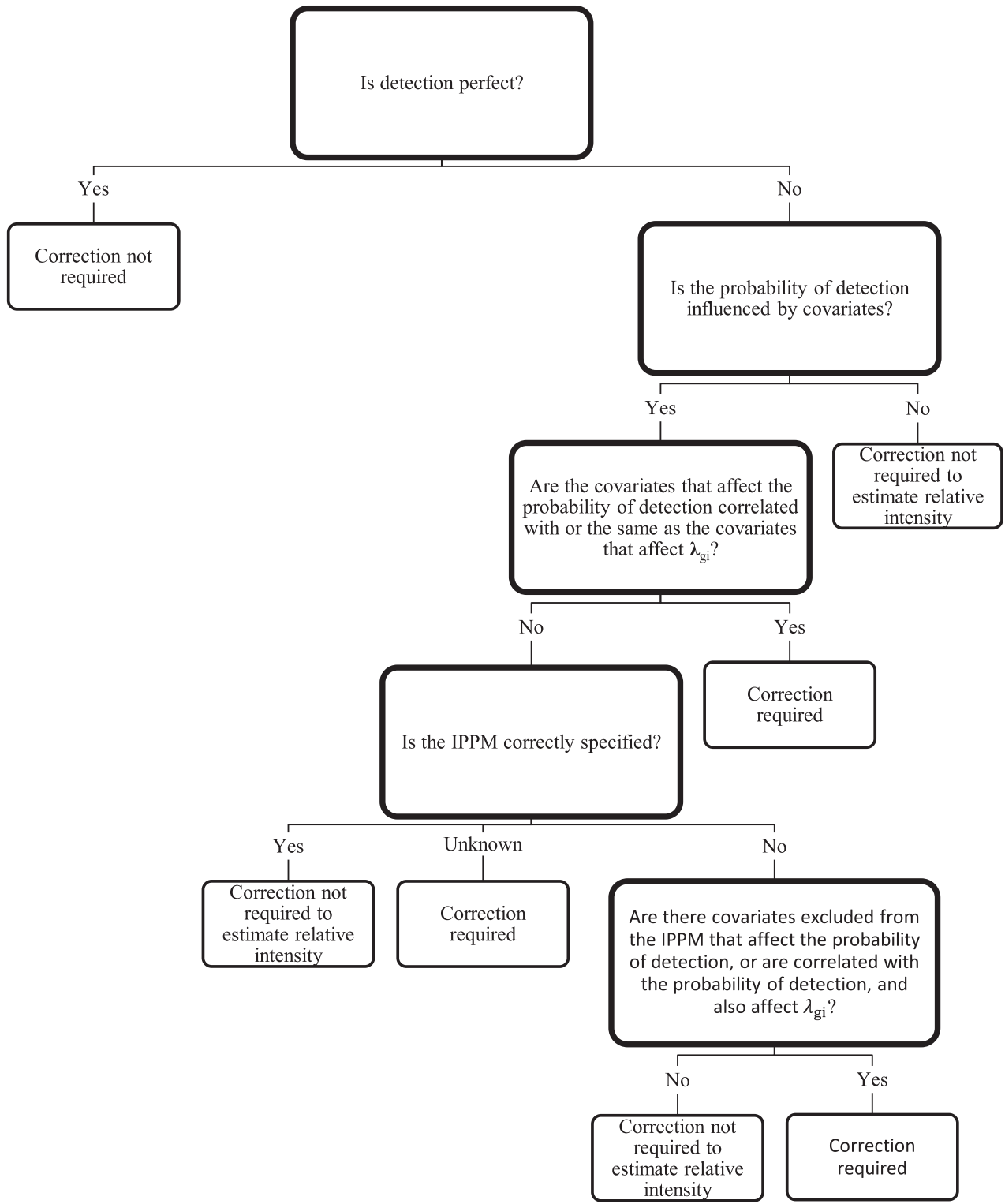

**Figure 1.** Decision tree used to determine whether correcting for nondetection sampling bias is required when analyzing presence-only data using an inhomogeneous Poisson point process model (IPPM).

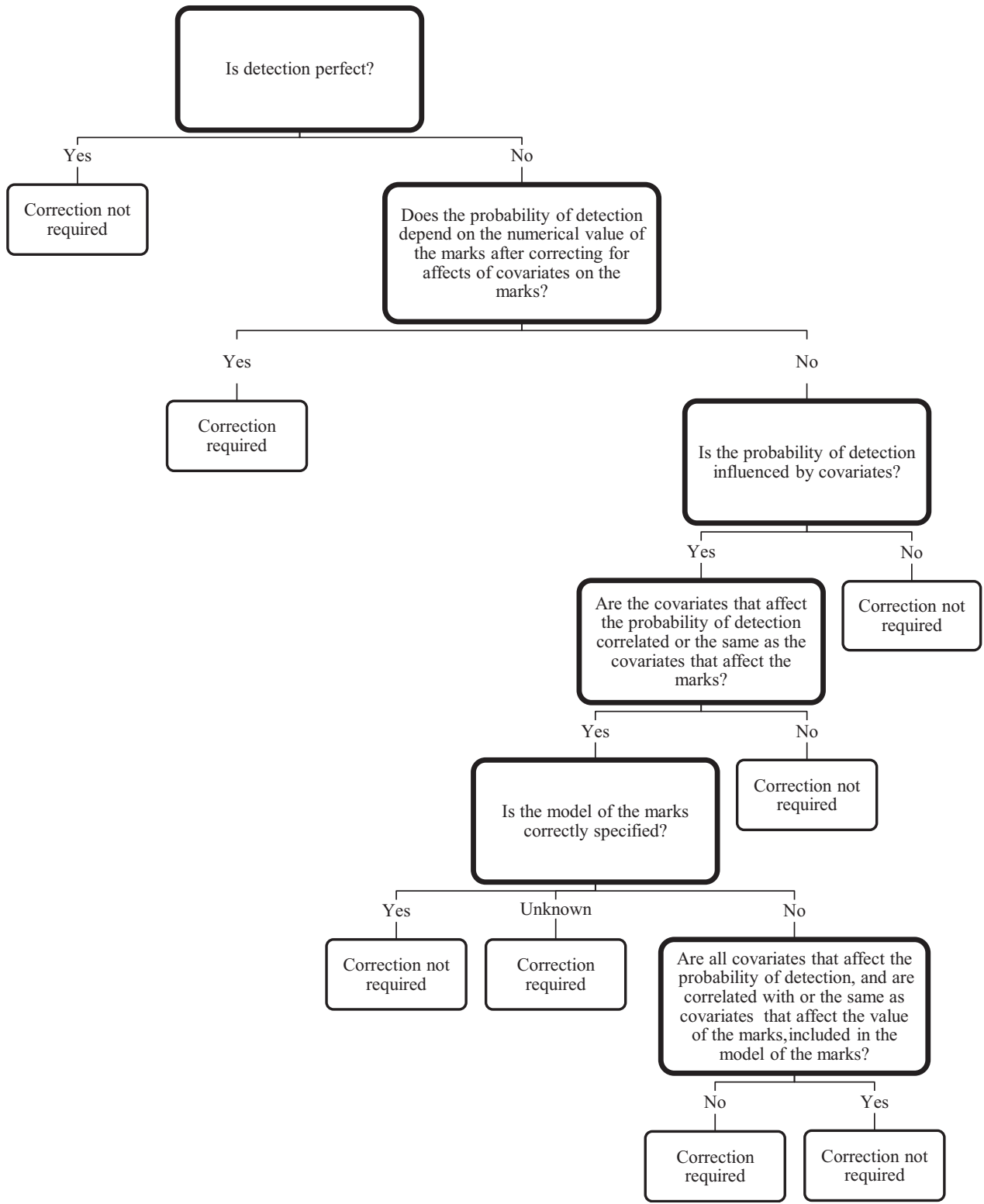

**Figure 2.** Decision tree used to determine whether correcting for nondetection sampling bias is required when analyzing marks (e.g., group sizes) associated with presence-only data.

For presence-only data, correcting for nondetection is the same as correcting for missing data; therefore, we used methods to correct for NIM data in our study. To correct for NIM data, estimates of $p_{det}$ must be obtained from auxiliary data (henceforth referred to as the detection data set) as there is no information in presence-only data about the detection process (Rubin 1976; Little and Rubin 2002). To correct for NIM data, the inverse of $p_{det}$ is used to weight the log-likelihood of the IPPM and zero-truncated GLM (Little and Rubin 2002). Correcting for nondetection by weighting the log-likelihood is attractive because the analysis can be carried out in standard software that allows weights to be specified (see Appendix S1 for annotated R code).

Although weighting the log-likelihood corrects the bias in the coefficient estimates and predictions of $\lambda_{gi}$ and $\lambda_{gs}$, obtaining meaningful measures of uncertainty such as standard errors (SE), confidence intervals (CI), and prediction intervals that incorporate the uncertainty in the detection process requires additional effort in the form of implementing a two-phase bootstrapping algorithm. We implemented a two-phase, nonparametric bootstrap algorithm which uses the detection data set to obtain estimates of $p_{det}$ and then fits the marked IPPM using the estimates of $p_{det}$ to correct for nondetection sampling bias. We present the algorithm here:

(1) Draw a bootstrap sample from the detection data set.
(2) Fit an appropriate model to the detection data set.
(3) Draw a bootstrap sample from the presence-only data that includes group size marks.
(4) Estimate $p_{det}$ for each location for the bootstrap sample in step 3 using the fitted model from step 2.
(5) Fit an IPPM that weights the log-likelihood function using $1/\widehat{p_{det}}$ and save coefficient estimates or predicted values of $\lambda_{gi}$.
(6) Fit a model to group size that weights the log-likelihood function using $1/\widehat{p_{det}}$ and save coefficient estimates or predicted values of $\lambda_{gs}$.
(7) Repeat steps 1–6 to obtain $b$ bootstrap samples.

The CI and SE can be calculated from the empirical distributions; however, many other summaries of the empirical distributions (e.g., mean) may be of interest (Efron and Tibshirani 1994). An annotated example with R code implementing the two-phase nonparametric bootstrapping algorithm for the IPPM and zero-truncated GLM is available in Appendix S1.

The use of weighted log-likelihoods to correct for bias has a long history for NIM data (Little and Rubin 2002) and has been used successfully to account for NIM data when GPS collars fail to record animal use locations in habitat selection studies (Frair et al. 2004). Although weighting provides an automatic procedure to reduce bias in parameter estimates and predictions from the IPPM and zero-truncated GLM when detection bias results in NIM data, weighting results in an increase in variance of the estimands. The increased variance maybe undesirably large and thus correcting for nondetection should be viewed as a bias–variance tradeoff. In general, imprecise (i.e., due to small sample size) and highly variable (i.e., due to the effect of covariates) estimates of $1/p_{det}$ will result in highly variable estimands from the IPPM and zero-truncated GLM. For our simulation study, we estimated $p_{det}$ using logistic regression (see simulation study); however, methods such as regularization that result in coefficient shrinkage or trimming that result in less variable estimates of $1/p_{det}$ may result in a more desirable bias–variance tradeoff (Little and Rubin 2002; Hastie et al. 2009).

## Simulation study

We conducted a simulation study to assess the properties of our SDM. For our simulation study, the data-generating distributions corresponded to those of the IPPM and zero-truncated GLM. This allowed us to test our two-phase bootstrap algorithm and determine whether our algorithm performed well on simulated data where the true values were known. We simulated group presence-only data ($y_{pres}$) over a region with 1 million pixels using an inhomogeneous Poisson point process distribution with intensity function ($\lambda_{gi}$) that varied according to the linear predictor:

$$\log(\lambda_{gi}) = \alpha_0 + \alpha_1 z_{gi}, \quad (4)$$

where $\alpha_0$ was the intercept and $\alpha_1$ was the regression coefficient for the vector of covariates $z_{gi}$. At each presence location, group sizes ($y_{gs}$) were simulated using a zero-truncated Poisson distribution with a rate parameter ($\lambda_{gs}$) that varied according to the linear predictor:

$$\log(\lambda_{gs}) = \gamma_0 + \gamma_1 z_{gs}, \quad (5)$$

where $\gamma_0$ was the intercept and $\gamma_1$ was the regression coefficient for the vector of covariates $z_{gs}$. Detection of each group ($y_{det}$) was simulated using a Bernoulli distribution, where a realized value of one represented detection and a value of zero represented nondetection. The probability of detection ($p_{det}$) varied according to the linear predictor:

$$\text{logit}(p_{det}) = \theta_0 + \theta_1 z_{det} + \theta_2 s(y_{gs}), \quad (6)$$

where $\theta_0$ was the intercept, $\theta_1$ was the coefficient for the vector of covariates $z_{det}$, and $\theta_2$ was the coefficient for the scaled and centered effect of group size ($s(y_{gs})$).

The entire simulated data set could be represented by the vectors: $y_{pres}$, $y_{gs}$, $y_{det}$, $z_{gi}$, $z_{gs}$, and $z_{det}$. The observed presence-only data set was comprised of groups that were detected (i.e., $y_{det} = 1$). The auxiliary data used to estimate and correct for detection bias were obtained by taking a random sample without replacement from the full simulated data set (detected and nondetected). Logistic regression was used to estimate $p_{det}$ using the auxiliary data set assuming the linear predictor in equation (6).

We simulated data from the worst-case scenario: low detection in habitat with a high intensity of abundance (i.e., more and larger groups) and where the covariate that affects the intensity of abundance is the same as the covariate that affects detection. We simulated the covariates from a single standard normal distribution so the covariates of group intensity, group size, and detection were the same (i.e., $z_{gi} = z_{gs} = z_{det}$). The covariate parameter for the inhomogeneous Poisson point process distribution was fixed at $\alpha_1 = 1$. We evaluated two sample sizes by setting the intercept ($\alpha_0$) to 7.0 for the small sample size and 8.5 for the large sample size. We conducted 1000 simulations for each sample size and estimated the parameters of the IPPM using infinitely weighted logistic regression with 1000 Monte Carlo integration points and weights of 10000 (Fithian and Hastie 2013). The parameters for the zero-truncated Poisson distribution used to simulate group size were $\gamma_0 = 1$ and $\gamma_1 = 0.5$. The parameters for the Bernoulli distribution used to simulate the detection process for groups were $\theta_0 = -2$, $\theta_1 = -1$, and $\theta_2 = 0.5$, so that detection decreased with the habitat covariate and increased with group size. We randomly sampled 20% of the full data set to obtain our auxiliary detection data and estimated $p_{det}$ using logistic regression. Extremely low values in $p_{det}$ in the small sample size case resulted in convergence issues for steps five and six in our two-phase bootstrap algorithm, so we trimmed $\widehat{p_{det}}$ by replacing values in $\widehat{p_{det}} \leq 0.01$ with 0.01. Although trimming $\widehat{p_{det}}$ could result in biased coefficient estimates, it improved convergence and greatly reduced the variance of parameter estimates from the IPPM and zero-truncated GLM with a minimal increase in bias in our simulations. For each simulation, we used $b = 1000$ bootstrap samples to estimate statistics from the empirical distributions.

We evaluated the results from our simulations by plotting the mean of the empirical distributions of $\alpha_1$, $\gamma_1$, and $e^{\alpha_1 + \gamma_1}$ from each simulation and compared it to the known value. For management purposes, $\widehat{\alpha_1}$, $\widehat{\gamma_1}$, and $\widehat{e^{\alpha_1 z_{gi} + \gamma_1 z_{gs}}}$ would likely be the parameters of most interest. The $e^{\alpha_1 z_{gi} + \gamma_1 z_{gs}}$ describes the relationship between the relative intensity of abundance and the environmental covariates, which could be used to compare two different points or areas to evaluate the relative conservation value, in terms of expected relative abundance, of each area for the species of interest.

Our two-phase bootstrap algorithm was complicated and involved several connected models. In theory, our algorithm should produce estimates with good frequentist properties, and to verify this, we calculated the coverage probability of the 95% CIs obtained from the 2.5th and 97.5th percentiles of the empirical distributions of $\alpha_1$, $\gamma_1$, and $e^{\alpha_1 + \gamma_1}$. To assess the effects of sample size, we calculated the scaled length (length/effect size) of the 95% CIs for $\widehat{\alpha_1}$, $\widehat{\gamma_1}$, and $\widehat{e^{\alpha_1 + \gamma_1}}$ and compared the small and large sample sizes. We plotted CI coverage probability against scaled CI length to allow for simultaneous evaluation of coverage probability and sample size.

We evaluated the properties of our statistical methods by comparing the results from the five scenarios for each sample size: (1) $p_{det}$ was estimated and used to correct for detection bias; (2) $p_{det}$ was estimated but the detection model was misspecified due to unknown group size; (3) $p_{det}$ was known; (4) an unbiased sample of group locations and sizes (i.e., detection was perfect) was analyzed; and (5) detection bias was ignored. For studies using our methods, group size may be unknown in some of the auxiliary detection data (e.g., nondetected groups in a telemetry study; see discussion). Because of this, we evaluated our models ignoring the effect of group size (scenario 4) and estimated the parameters in our detection model with the misspecified linear predictor:

$$\text{logit}(p_{det}) = \theta_0 + \theta_1 z_{det}. \quad (7)$$

Misspecification of the detection model could result in biased estimates of $p_{det}$, which, in turn, would result in biased estimates of $\alpha_1$, $\gamma_1$, and $e^{\alpha_1+\gamma_1}$. If the estimated $\boldsymbol{p}_{\text{det}}$ does not depend on group size or if group size was not available, there is no need to provide weights $(1/\widehat{\boldsymbol{p}_{\text{det}}})$ in step six of our estimation algorithm because the correction is equivalent to assuming that missing group size marks were MAR.

We compared estimates of $\alpha_1$, $\gamma_1$, and $e^{\alpha_1+\gamma_1}$ from simulations of all five scenarios. We designed the comparison between the parameter estimates when $\boldsymbol{p}_{\text{det}}$ was known (scenario 3) to those when $\boldsymbol{p}_{\text{det}}$ was estimated (scenarios 1 and 2) to show the increase in variance due to uncertainty in $\widehat{\boldsymbol{p}_{\text{det}}}$. We designed the comparison between parameters estimates from the unbiased sample (scenario 4) and when $\boldsymbol{p}_{\text{det}}$ was known (scenario 3) to illustrate the increased variance of estimated parameters due to weighting the log-likelihood. Finally, we compared estimates from scenarios 1–4 to estimates from data when detection was ignored and the data were assumed to have been derived from an unbiased sampling effort (scenario 5).

## Results

The average number of presence-only groups in each simulation was 1809.19 (SD = 41.52) and 8098.87 (SD = 88.61) for the small and large sample size, respectively. The average probability of group detection was 0.06 (SD = 0.05) and resulted in average sample sizes of 108.12 (SD = 10.42) and 483.44 (SD = 21.80) presence-only locations. The auxiliary detection data had an average sample size of 362.84 (SD = 8.30) and 1619.77 (SD = 17.72) with an average of 21.60 (SD = 4.51) and 96.70 (SD = 9.54) detections. The average group size was 5.18 (SD = 3.46) for all groups and 4.49 (SD = 3.19) for all detected groups. The bootstrap algorithm converged in all of our simulations.

For the simulation that included small sample size, a known group size in the auxiliary detection data, and when $\boldsymbol{p}_{\text{det}}$ was estimated (scenario 1), $\widehat{\alpha}_1$, $\widehat{\gamma}_1$, and $\widehat{e^{\alpha_1+\gamma_1}}$ had minimal bias (−0.014, 0.020, 0.240) and small variance (0.036, 0.003, 0.909; Fig. 1). When group size was unknown in the auxiliary detection data (scenario 2), $\widehat{\alpha}_1$ and $\widehat{e^{\alpha_1+\gamma_1}}$ were generally more biased (0.046, 0.534) and variable (0.050, 1.426), but $\widehat{\gamma}_1$ had the same bias (0.005) and variance (0.002) as when detection was ignored because the correction was equivalent to assuming the group size marks were MAR, and was therefore not applied. When $\boldsymbol{p}_{\text{det}}$ was known (scenario 3), $\widehat{\alpha}_1$, $\widehat{\gamma}_1$, and $\widehat{e^{\alpha_1+\gamma_1}}$ were less biased (0.001, 0.009, 0.097) and less variable (0.024, 0.004, 0.495) than when $\boldsymbol{p}_{\text{det}}$ was estimated with known group size (scenario 1). The $\widehat{\alpha}_1$, $\widehat{\gamma}_1$, and $\widehat{e^{\alpha_1+\gamma_1}}$ had the lowest combination of bias (0.001, 0.000, 0.085) and variance (0.016, 0.002, 0.410) when an unbiased sample of presence-only locations was used (scenario 4). Finally, when detection was ignored (scenario 5), $\widehat{\alpha}_1$, $\widehat{\gamma}_1$, and $\widehat{e^{\alpha_1+\gamma_1}}$ were highly biased (−0.646, 0.005, −2.105) with low variance (0.011, 0.002, 0.075). Our results were nearly identical for the larger sample size, except the variances decreased when sample size was increased (Fig. 3).

Coverage probabilities of 95% CIs for $\widehat{\alpha}_1$, $\widehat{\gamma}_1$, and $\widehat{e^{\alpha_1+\gamma_1}}$ were close to 0.95 for the scenario when group size was known in the auxiliary detection data and detection was estimated (scenario 1). When group size was unknown and detection was estimated (scenario 2), coverage probabilities for $\widehat{\alpha}_1$, $\widehat{\gamma}_1$, and $\widehat{e^{\alpha_1+\gamma_1}}$ were close to 0.95 for the small sample size, but slightly less than the nominal level for the larger sample size. As expected, standardized 95% CI lengths decreased as sample size increased (Fig. 4). We did not evaluate the coverage probabilities or effects of sample size for scenarios 3–5, because they did not require implementation of the bootstrap algorithm.

## Discussion

The equivalence of nondetection sampling bias and NIM data has profound implications for SDMs using presence-only data because the missing data mechanism (i.e., MCAR, MAR, and NIM) cannot be determined from the data at hand (Rubin 1976; Little and Rubin 2002). As a result, the effects on nondetection sampling bias cannot be determined from presence-only data without auxiliary detection data. When nondetection results in NIM data and is ignored in the analysis, the realized, rather than the true, distribution of abundance is estimated (Kéry 2011). The true distribution of abundance is not identifiable from presence-only data without assuming nondetection results in MCAR data. As a result, auxiliary detection data are required to determine whether the coefficient estimates of environmental features are related to the true distribution of abundance, the detection process, or both. This result has strong implications for analyses using SDMs with presence-only data because if the detection process results in NIM data and is ignored, the SDM cannot separate environmental features affecting the distribution of species' abundance from those affecting detection of the species.

At a minimum, considering the implications of nondetection and exploring corrective measures should be an essential part of analyses using presence-only data. However, the crux of the exploration and correction for the effects of nondetection is obtaining auxiliary data to assess the detection process. We suspect that for most opportunistic presence-only data sets, especially for mobile species, these auxiliary data do not exist. For the whooping crane records that motivated the development of methods in this study, we are pursuing and recommend for other mobile species, two sources of potential data: telemetry and expert elicitation. If a proportion of the study population could be telemetered, the presence-only records could be matched to telemetered animals. Presence-only records that occur at the same place and time as a telemetered animal is detections (i.e., 1s); non-detections are telemetry locations of groups not detected (i.e., 0s). The data could be analyzed, as in our simulation study, with logistic regression. Based on results from our simulation study, the number of detections required may be relatively small (e.g., ~20) to result in adequate correction of nondetection sampling bias. Use of telemetry data, however, is based on an implicit assumption that the detection model, and data are transportable. Transportability of the detection model and data requires an assumption that the detection process for the telemetered animals during the time period of the telemetry study was similar to that of the presence-only records. This assumption, however, may be impossible to verify. Because of this, obtaining auxiliary detection data from telemetered animals will not be useful for the majority of studies that analyze historical presence-only records. An alternative source of data is experts. Expert elicitation may be the only feasible means of obtaining the auxiliary data necessary to explore the effects of and correct for nondetection sample bias for historical presence-only records. Expert elicitation is well developed for ecological studies (Martin et al. 2012; Perera et al. 2012) and has been used for studies with NIM data (White et al. 2007; Jackson et al. 2010; Mason et al. 2012).

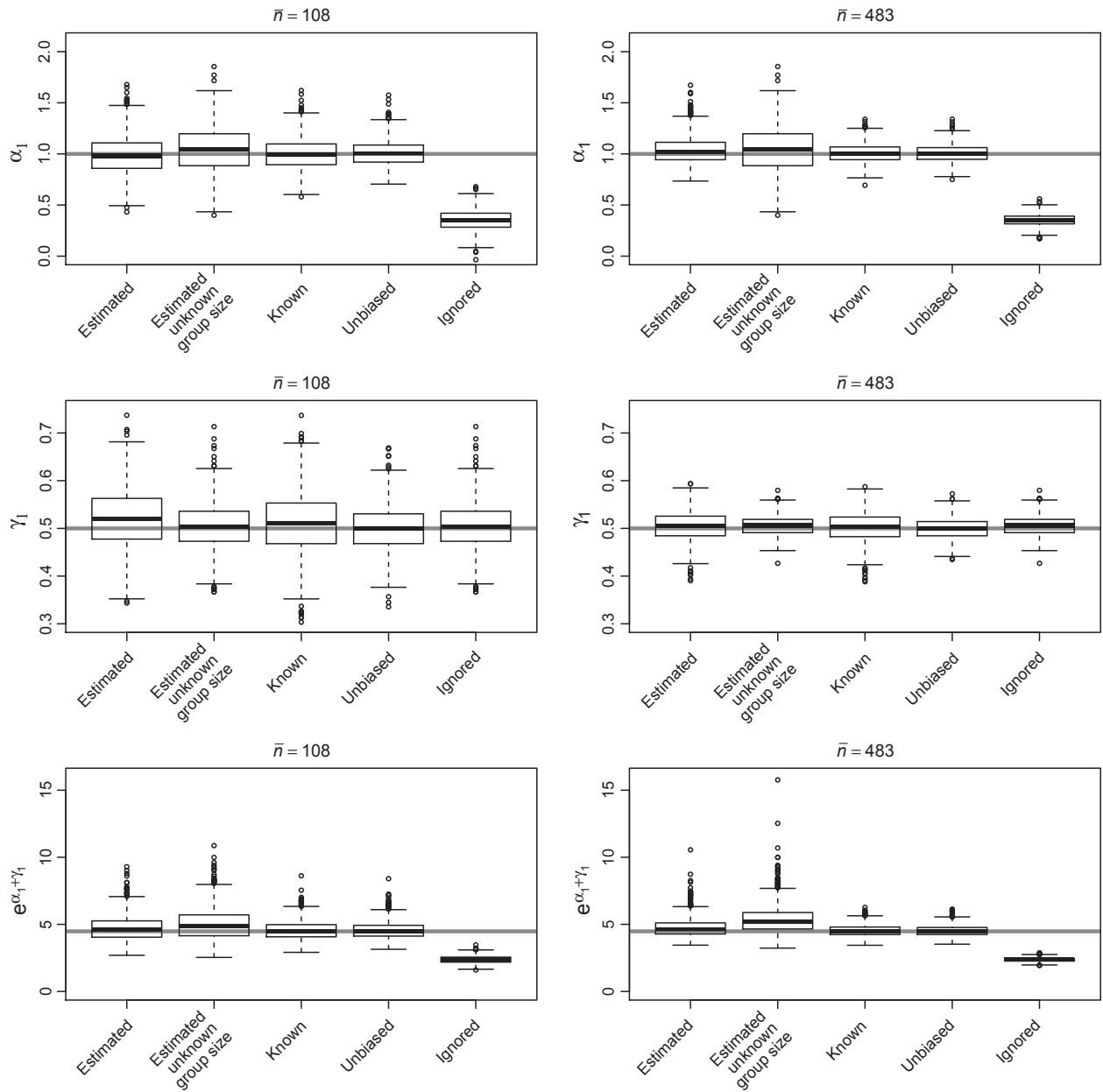

**Figure 3.** Regression coefficient estimates from simulated data using an IPPM ($\alpha_1$) and zero-truncated GLM ($\gamma_1$) to describe how the relative intensity of group abundance and expected group size varied due the respective covariate. The $e^{\alpha_1+\gamma_1}$ was a derived parameter that described the relative intensity of abundance. The five scenarios shown include scenarios in which $p_{det}$ was estimated and used to correct for detection bias (Estimated; scenario 1), $p_{det}$ was estimated but the detection model was misspecified due to unknown group size (Estimated unknown group size; scenario 2), $p_{det}$ was known (Known; scenario 3), an unbiased sample of group locations was analyzed (Unbiased; scenario 4), and detection bias was ignored (Ignored; scenario 5). Each box and whisker corresponds to parameters estimates obtained from 1000 simulated data replicates, and the grey lines represent the true value. We evaluated two parameterizations that resulted in observed average sample sizes of 108 and 483.

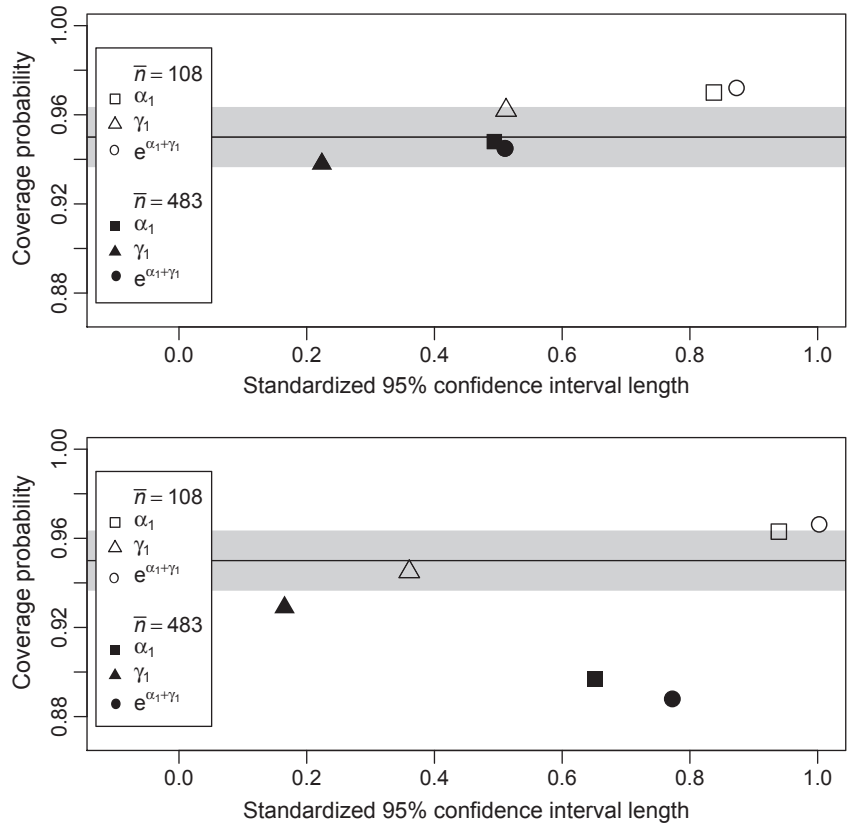

**Figure 4.** Coverage probability of 95% confidence intervals (CI) plotted against the standardized 95% CI length from simulated data using the IPPM ($\alpha_1$) and zero-truncated GLM ($\gamma_1$) to describe how the relative intensity of group abundance and expected group size varied due to the respective covariate. The $e^{\alpha_1+\gamma_1}$ was a derived parameter that described the relative intensity of abundance. We evaluated two sets of parameters that resulted in observed average sample sizes of 108 and 483. The two scenarios shown include when $p_{det}$ was estimated and used to correct for detection bias (upper panel; scenario 1) and when $p_{det}$ was estimated, but the detection model was misspecified due to unknown group size (lower panel; scenario 2). Horizontal lines were placed at 95% coverage probabilities with 95% CI coverage based on a normal approximation (grey shaded areas).

Studies documenting the relationship between environmental features and a species' distribution of abundance must consider the grouping behavior of individuals. For example, the location of birds within a flock could be highly, if not, perfectly correlated. Because of this behavior, the standard IPPM is appropriate to model the distribution of group abundance. We illustrated how to model the distribution of species' abundance by treating group sizes as marks. Based on our theoretical and numerical simulation results, the IPPM and zero-truncated GLM provide a framework to combine models of group intensity and size. The strength of our framework is that it accounted for the extreme correlation between individuals in a group and allows us to model group intensity and group size independently.

We explored the effects of nondetection bias, and our results for the marked IPPM were comparable to other studies (Dorazio 2012). By framing the nondetection sampling bias as a missing data mechanism, we were able to provide a unified framework that could be applied to both group locations and group size marks in addition to utilizing bias correction methods that were developed for missing data. The results from our numerical simulations were encouraging. When the data-generating mechanisms corresponded to the models used in the analysis, coefficients obtained using the two-phase bootstrap algorithm had good frequentist properties. The parameter estimates were centered on the true value, and the CIs had near nominal coverage (Figs. 3, 4).

We observed an increase in variance of the corrected IPPM and zero-truncated GLM parameter estimates in the results of our simulation analysis. This will likely occur whenever one corrects for nondetection or NIM data (Fig. 3). The general conclusions about the benefits of correcting for NIM data include the following: (1) the amount of bias, and hence bias correction needed, will vary depending on the data set, (2) the increase in variance could offset any beneficial reduction in bias, and (3) bias correction should not be automatically applied and assumed to provide reliable results due to point number two (Little and Rubin 2002). We feel these conclusions are equally relevant when correcting SDMs for nondetection bias. For example, to obtain asymptotically unbiased estimates of the IPPM and zero-truncated GLM coefficient estimates, we needed unbiased estimates of $p_{det}$ from logistic regression. For our numerical simulation (with small sample size), this resulted in convergence issues and highly variable estimates of coefficients of environmental covariates and associated CIs that were orders of magnitude wider than those obtained when the bias was ignored. Because of this, we trimmed the estimates in $p_{det}$ as described in our methods. Trimming results in asymptotic bias, but for our realized sample sizes, the bias was minimal and the reduction in variance was large. Development of data driven methods for trimming $p_{det}$ when correcting for non-detection bias in SDM is needed (Elliott 2007).

Correcting for nondetection is difficult, but these difficulties are not limited to presence-only data. For example, correction of nondetection in species occupancy models using presence–absence data where nondetection results in false negatives can be exceedingly difficult (Welsh et al. 2013). Our methods can only be used if adequate auxiliary data are available; however, practitioners must consider the well-known bias–variance trade off. Alternatively, the detection process could be ignored, and a sensitivity analysis could be conducted (White et al. 2007; Johnson and Gillingham 2008; Jackson et al. 2010; Mason et al. 2012).

# Acknowledgments

We thank M. Sliwinski and two anonymous reviewers for their review and comments on this manuscript. This research was supported by funding from the Platte River Recovery Implementation Program and the National Science Foundation Integrative Graduate Education and Research Traineeship (NSF-DGE-0903469).

# Conflict of Interest

None declared.

## Supporting Information

Additional Supporting Information may be found in the online version of this article:

**Appendix S1.** Annotated R code used to implement marked Poisson point process species distribution model and two-phase nonparametric bootstrap algorithm.